\documentclass[journal]{IEEEtran}
\ifCLASSINFOpdf
\else
\fi
\usepackage{amsmath}

\usepackage{esint}
\usepackage{subfigure}

\usepackage{cite}
\usepackage{graphicx}
\begin{document}
%
\title{A Subwavelength-Laser-Driven Transmitting Optical Nanoantenna for Wireless Communications}

\author{Amer~Abu~Arisheh,
        Said~Mikki, and
        Nihad~Dib.

\thanks{A. Abu~Arisheh and N. Dib are with the ECE Department of Jordan University of Science \& Technology, Irbid, Jordan. S. Mikki is with the Department of ECECS, University of New Haven, West Haven, CT, USA.}}

\markboth{}%
{Shell \MakeLowercase{\textit{et al.}}: Bare Demo of IEEEtran.cls for IEEE Journals}
%



\maketitle

\begin{abstract}
Nanoantennas are efficient devices exhibiting large confined electric field enhancements.
So far, they have been extensively researched mainly in the receiving mode, which means that the illuminating field is essentially a plane wave. In this paper, we consider the problem of designing an efficient and highly directive
transmitting Nanoantenna where the system is energized by a non-plane wave field, a subwavelength laser excitation. Including short-wavelength components allowed us to achieve a 200-nm spot radius, which is a quarter of its incident wavelength (800 nm). Near- and far-field antenna quantities are introduced and calculated using an efficient full-wave multiphysics solver. A nano-scale optical antenna is then presented with optimized dimensions and material settings. Various design curves and insights are also discussed in connection with how issues such as how to define efficiency and determine whether the system is radiating properly. 
\end{abstract}


%
\IEEEpeerreviewmaketitle

\section{Introduction}

\IEEEPARstart{R}{esearch} on nano-antennas for visible and infrared radiation is an emerging field with novel applications in many areas such as biomedical imaging, near-field nano-optics, quantum communications, and optical signal processing \cite{novotny2012principles}. The importance of nanoantennas comes from their ability to provide a substrate for observables and potential useful interactions between light and structures that are much smaller than the diffraction limit. In RF antennas, the conventional role of the antenna device is to link RF waves in space to transmission lines that are comparable with the wavelength \cite{balanis2012}. Analogously, the nanoantenna allows light localization at a subwavelength scale by establishing light-matter interactions involving objects whose size is below the diffraction limit \cite{csendur2009near}. In fact, the analogy is not perfect since there are two main differences between RF antennas and nanoantennas. First, skin depth at optical frequencies is appreciable, while at RF frequencies, metal can be safely treated as a perfect electric conductor (PEC) \cite{novotny2011near}. Second, surface plasmon resonances appear when nano-structures are illuminated with an optical wave \cite{csendur2009near}.
However, similar to RF devices, nanoantennas come out in a diverse variety of geometric types such as dipole, Yagi-Uda, bow-tie, etc, which involve geometric resonance processes as well as plasmonic resonances \cite{muehlschlegel2005resonant,maksymov2012optical,yousif2012modeling}. Nanoantennas have been extensively designed and studied in the case of plane wave excitation such as in \cite{zainud2016single,el2017tapered,razzari2011extremely,aizpurua2005optical,liu2012comparison}. Focus on using the plane wave excitation techniques typically implies that in the mainstream literature it is often the \textit{receiving mode} operation that is considered. On the other hand, publications dealing with transmitting mode investigate nanoantennas that are typically excited by a nanosized optical-frequency generator \cite{see2018photoluminescence} such as a quantum emitter \cite{curto2010unidirectional}. Focused lasers can also be deployed for excitation because they are able to concentrate the EM field within a domain smaller than the diffraction limit. This allows the laser field to illuminate parts of the nanoantenna even if the wavelength of the laser is in the optical regime. 

The idea of exciting a nanoantenna with a focused laser was investigated by some workers in the past such as  \cite{see2018photoluminescence,csendur2009near,ghenuche2008spectroscopic}, though a fully-fledged Tx nanoantenna using focused laser wasn't numerically simulated as a part of a nanoscale wireless communication system. 
In \cite{csendur2009near}, a nanodipole antenna was designed for functional plasmonic applications. The authors considered the power transmitted to a sample placed in front of the nanodipole while the nanodipole is illuminated by a focused laser. The focusing was achieved using a lens. A model was developed to compute the near-field power transmission, and Finite-Element Method (FEM) simulations were performed to study the effect of the nanodipole antenna dimensions and choose optimum dimensions. In \cite{ghenuche2008spectroscopic}, single and coupled gold nano-wires were illuminated using a focused laser beam. Two-Photon Induced Luminescence (TPL) microscopy was utilized to assess the salient features of 
resonance phenomena in gold nanoantenna such as central wavelength and bandwidth. In the TPL framework and setting, the antenna illumination was basically a focused field with a 700-780 nm laser beam realized in the lab by means of immersion oil (100X objective) with 1.25 numerical aperture. The illumination spot was around 350 nm. In \cite{see2018photoluminescence}, broadband
transmitting directional nanoantennas were experimentally realized using plasmon-modulated photoluminescence (PMPL) as an effective optical driving source. One-photon photoluminescence from gold nanostructures was excited by a circularly polarized 532 nm CW laser. The excitation laser was focused onto nanoantenna from the substrate side by an oil objective (PlanApo 60X Oil N.A. = 1.42, Olympus).

In this paper, however, we consider a different method to excite a transmitting nanoantenna, which is via illumination by a subwavelength laser. Subwavelength laser is a method to generate coherent optical fields at the nano-scale (beyond diffraction limit) \cite{oulton,amersaidnihad}. It is a promising field in experimental optics and optical electronic devices; e.g., some researchers have reported the experimental demonstration of nanometer-scale lasers generating a 100 times smaller than the diffraction limit \cite{oulton}. 
To the best of our knowledge, illuminating a nanoantenna by a subwavelength laser has not been discussed in the literature so far, aside from our recent brief report \cite{amersaidnihad}. In this paper, a detailed and full investigation of how to model and design transmitting nanoantenna using a subwavelength laser excitation method is provided. The nanoantenna is simulated and optimized for a high directivity and efficiency using the multiphysics solver COMSOL \cite{comsol}. 
The paper is organized as follows: In Section \ref{sec:Principal Models and Design Methodology}, the simulated nanoantenna modelling stages are explained involving geometry, material, physics, mesh and study in addition to near-field and far-field definitions that were manually entered in the post-processing stage. In Section \ref{sec:Computational Analysis and Design Examples}, the results of the parametric study are discussed and the optimum transmitting nanoantenna is determined. Also, a nanoantenna communication system is considered from the perspective of polarization diversity where corresponding optimum  dimensions are determined. Section \ref{sec:Conclusion and Future Work} concludes the work and proposes some ideas and potential paths of improvement to be explored in the future.

\section{Principal Models and Design Methodology}
\label{sec:Principal Models and Design Methodology}

In what follows, we deploy COMSOL Multiphysics to model the geometry, material, physics and mesh of the proposed subwavelength optical nanoantenna model. Also, basic design methodology and postprocessing expressions are defined and discussed in detail. This computational package has already been used by some researchers in the past to model nanoantennas, e.g., see \cite{yushanov2013mie,afridi2016beam,nafari2017modeling,garcia2011strong,el2017tapered,calderon2014bowtie,liu2008plasmonic}. However, here we focus on how COMSOL provides some tools that facilitate modeling nanoantennas using the wave-optics module for subwavelength illumination in transmitting mode. 

\subsection{Outline of the Subwavelength Computational Model}

We first review some of the existing optical models that are relevant to our problem here. The two main cases involve the `Optical Scattering Off of a Gold Nanosphere' and `Scatterer on Substrate' models \cite{comsol}. In the former, a gold nanosphere was illuminated by a plane wave in a scattering formulation. Heat losses, skin depth, and radiation pattern were considered among others. In the latter, a gold nanoparticle scatterer placed on a dielectric substrate was illuminated by a TE-polarized electromagnetic wave in a full-field formulation. The absorption cross sections and scattering cross sections were evaluated for different azimuthal and polar angles. Some other models like `Self-Focusing', `Fresnel Lens', `Focusing Lens' and `Second Harmonic Generation of a Gaussian Beam (Wave Optics)' \cite{comsol} involve simulating laser beams focused by a lens or by Self-Focusing. However, in the `Nanorods' model \cite{comsol}, the nonparaxial Gaussian beam module, which will be extensively deployed in this work, was used to illuminate an array of thin rods. The Gaussian wave was generated using the nonparaxial Gaussian formula without breaking the diffraction limit, i.e. the spot radius is set to be equal to $\lambda$, not below it. 



\begin{figure}
\centering
\includegraphics[width=0.3\textwidth]{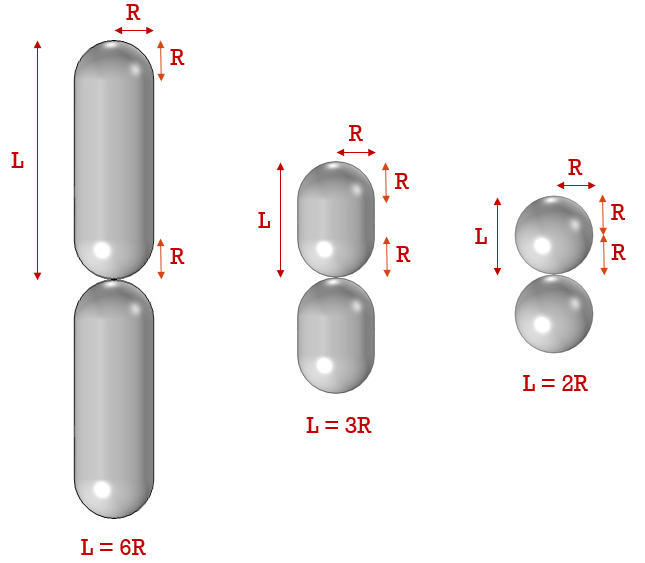}
\caption{Nanoantenna geometry for different ratios between arm Length $L$ and arm Radius $R.$ The gap length (not shown) between the two nanoparticles (nanodipole arms) is $g$.}
\label{Fig:extending}
\end{figure}

The geometries considered for our proposed nanoantenna are two-arm nanodipole and two nanospheres that are separated by a gap (see Fig. \ref{Fig:extending}.) The ends of the nanodipole arms are not sharp but intentionally made
smooth and rounded for several reasons. First, realistic nanorods do not have sharp ends; our selected geometry is closer to the actual forms taking shape in real fabricated prototypes \cite{ghenuche2008spectroscopic,eislersimulation}. Second, in sharp-corner nanodipole, the local field would not be maximal at the system center, while in our design it can exhibit a maximum at the middle of the gap between the two arms. This implies that the local field intensity enhancement factor (see below) can be defined unambiguously. Third, the rounded corners will make the geometry general and extendable to two nanospheres separated by a gap and this occurs as a special case of the general case geometry (nanodipole). It occurs when the arm length $L$ is set to be equal to $2R$, twice the radius as shown in Fig. \ref{Fig:extending}, i.e. when the upper and lower hemispheres of each arm are joined together. Regarding the gap between the two arms (nanoparticles), it is significant in the optical regime and is expected to exhibit high electric field intensities. The gap size decides if the two arms are highly coupled or not. Consequently, the length of the nanodipole is equal to $2L+g$, where $g$ is the gap size. The size and shape of our nanodipole (system of two nanoparticles) affect the excitation of surface plasmonic waves and consequently the overall resonance structure of the antenna. The radius of the nanodipole is significant because the surface plasmon wave excited on the two sides may mutually interact \cite{burke1986surface} and this interaction is maximum at a specific radius as found in \cite{csendur2009near}.

Benchmarked optical materials were used in our design for easier comparison with other published reports. In our model, classical gold permittivities for a range of wavelengths were used \cite{johnson1972optical}. For simplicity, the nanodipole antenna is suspended in a homogeneous medium (air in our case). It is also desirable to study the nanoantennas supported by thin substrates but this will unnecessarily complicate the analysis due to increased computation time and the possibility of exciting new surface waves on the substrate itself. Since the main goal of this paper is to provide first feasibility analysis of the potential to design subwavelength antennas, we focus on antennas suspended in free space in what follows. 


Gaussian beams deliver the highest power density for a given fixed incident power.\footnote{For more information about paraxial laser module visit this article on COMSOL BLOG:
`Understanding the Paraxial Gaussian Beam Formula'.} The conventional laser used in the lab and simulations can be described using the paraxial Gaussian beam formula.\footnote{\textit{Paraxiality} means that the laser beam is mainly propagating along the optical axis \cite{novotny2012principles}.} To ensure that this formula is accurate, the spot size (waist radius) of the laser beam should be at least ten times the operating wavelength. On the other hand, when the spot radius is close to or smaller than the wavelength, the beam propagates with a higher angle to the focus. Therefore, the paraxial approximation breaks down and here comes the importance of the \textit{non}paraxial Gaussian
beam formula which describes Gaussian beams in general. Subwavelength laser excitation is achieved using the nonparaxial Gaussian beam formulation, which is based on plane wave expansion. It is an exact solution for the Helmholtz equation unlike the Gaussian beam which is a solution for the paraxial approximation of the
Helmholtz equation \footnote{For more information visit this article on COMSOL BLOG:
'The Nonparaxial Gaussian Beam Formula for Simulating Wave Optics'}. The paraxial approximation formula is not suitable for our Tx nanoantenna design as we need an optical light focused on a subregion of the nanoscale particles, while the paraxial approximation formula is not accurate when the spot radius is equal to or smaller
than the wavelength. On the other hand, the angular spectrum expansion uses plane waves to approximate a Gaussian beam. The total expansion is a solution to the Helmholtz equation because each
single plane wave mode is a solution and the Helmholtz equation under consideration is linear. Consequently, since this approach can efficiently model a tightly focused beam, we will use it to simulate a subwavelength excitation field interacting with the nanodipole antenna using the full-wave analysis capabilities of COMSOL package. 

\begin{figure}
\centering
\includegraphics[width=0.3\textwidth]{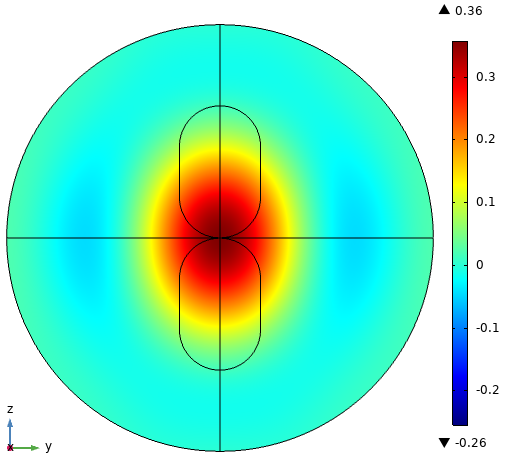}
\caption{\textit{z}-component of the background electric field (subwavelength laser excitation) with 800 nm
wavelength and 200 nm spot radius. The slice is taken at the plane \textit{x} = 0.}
\label{Fig:background1}
\end{figure}

In our model, the maximum transverse wave number in the plane wave spectrum expansion was limited to the free-space wavelength since evanescent waves are not included in the laser module. The number of plane waves used in our simulations is 338 plane waves. It can be increased to produce more accurate results but with the drawback of significantly increasing the computational burden. The
wavelength and spot radius were set to 800 nm and 200 nm, respectively. The laser illumination field was set to be $z$-polarized and propagating along the positive $x$ direction. The resultant subwavelength laser
(background field) is shown in Fig. \ref{Fig:background1}.
The maximum laser field is at the center and is around 0.36 V/m peak level. It is worth mentioning that for the nonparaxial module simulation, we used cubic discretization in order to obtain accurate results of the far-field radiation pattern, unlike many other COMSOL models where quadratic discretization is sufficient to get accurate results.

\begin{figure}
\centering
\includegraphics[width=0.3\textwidth]{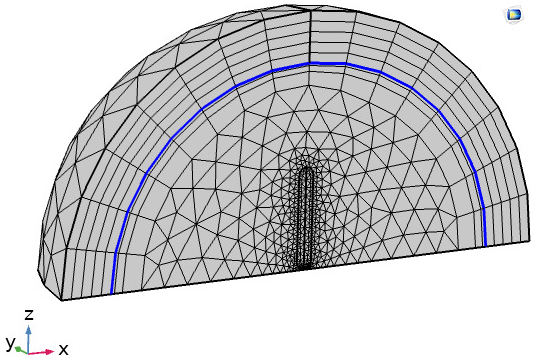}
\caption{The final mesh built on the simulated quarter of the model.}
\label{Fig:mesh}
\end{figure}

The symmetry in the nanodipole model is evident as the two arms are identical. Moreover, each arm possesses identical left and right sides with respect to the illumination field propagation vector, implying
that computational complexity can be reduced by only simulating a quarter of the original model while the rest can be inferred from the proper symmetry relations. The Perfectly Matched Layer (PML) is set at a distance of $\lambda/2+L+g/2$ from the center. The symmetry in the model is accounted for by defining a perfect magnetic conductor (PMC) in the $y= 0$ plane and a perfect electric conductor (PEC) in the $z = 0$ plane.
Grid settings are also a critical part. A reasonable number of discretization elements was selected by setting the maximum mesh element dimension to $\lambda/10$, which is sufficient for our parametric wavelength rage studies. This is important for reducing the computational burden required by electrically large geometry nanoantennas.
The far-field from antenna is computed in COMSOL using
 Stratton-Chu formula, which operates on the near-field information on the boundary between the air and the PML to produce the far-zone field. Three PML boundary layers were added
to enhance the far-field calculation. 
The final mesh built on the simulated quarter of the model is shown in Fig. \ref{Fig:mesh}.

\subsection{Basic Design Methodology}
\begin{figure}
\centering
\includegraphics[width=0.4\textwidth]{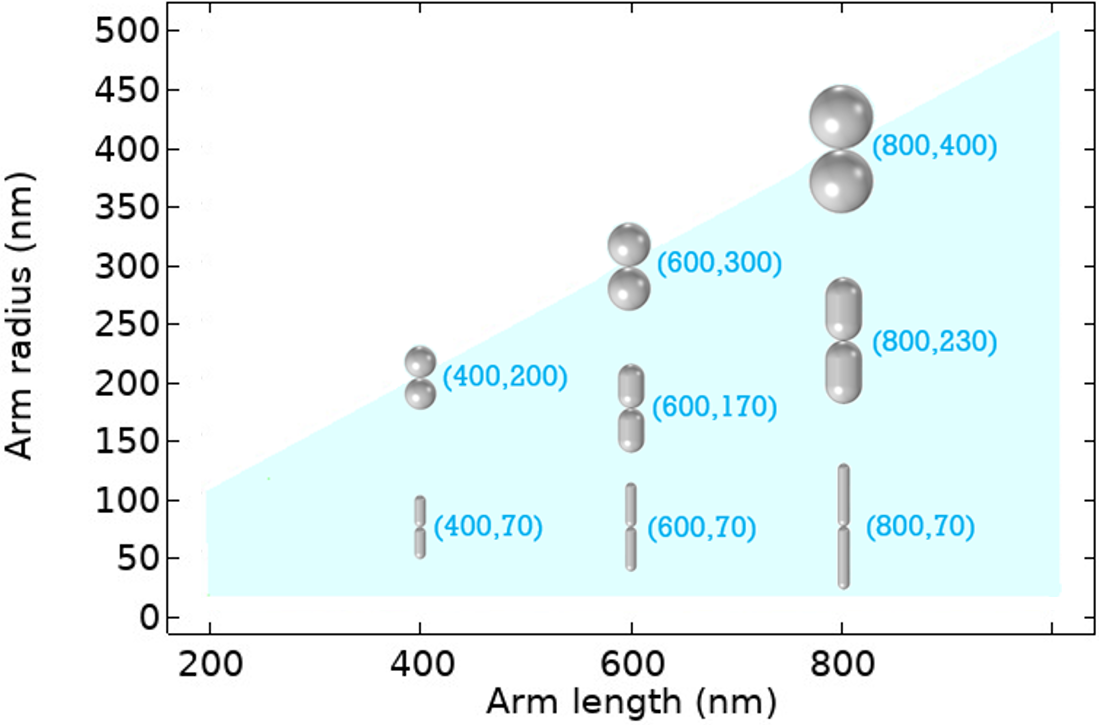}
\caption{Selected samples of the nanoantenna geometry at the indicated dimensions.}
\label{Fig:sense}
\end{figure}

The key idea in the design of all optical nanoantennas is to obtain (for a given optical material) the critical geometric dimensions and shapes that ensure the excitation of a strong surface plasmonic wave along the radiating dipole. In other words, a nanoantenna acts like a terminated optical waveguide or resonator where radiation occurs when a strong resonant current is excited due to the presence of a surface waveguide mode. In contrast to PEC dipoles operating at the lower microwave band, in most cases it is not possible to predict the critical resonance dimensions using analytical methods. Therefore, experimental methods (here full-wave numerical solution) are essential. Consequently, 
a parametric sweep was performed over the arm length from 200 nm to 1000 nm with 50 nm steps and over the radius from 10 nm to 500 nm with 20 nm steps. Not all combinations were taken into account since the case of $L<2R$ is excluded. In addition, the cases of perfect nano-spheres were added from 100 nm to 500 nm radii with 25 nm steps. The sweep over the length is basically equivalent to changing the wavelength of the incident field. Throughout this paper, we present most of the results in the form given in Fig. \ref{Fig:sense}, where some samples of the geometries taken in the sweep are given only in that Figure in order to induce a feel of the various design prototypes implied by such graphs hereafter. 

\subsection{Near-Field and Far-Field Analysis}
\label{sec:Near-Field and Far-Field Calculations}

\begin{figure}
\centering
\includegraphics[width=0.2\textwidth]{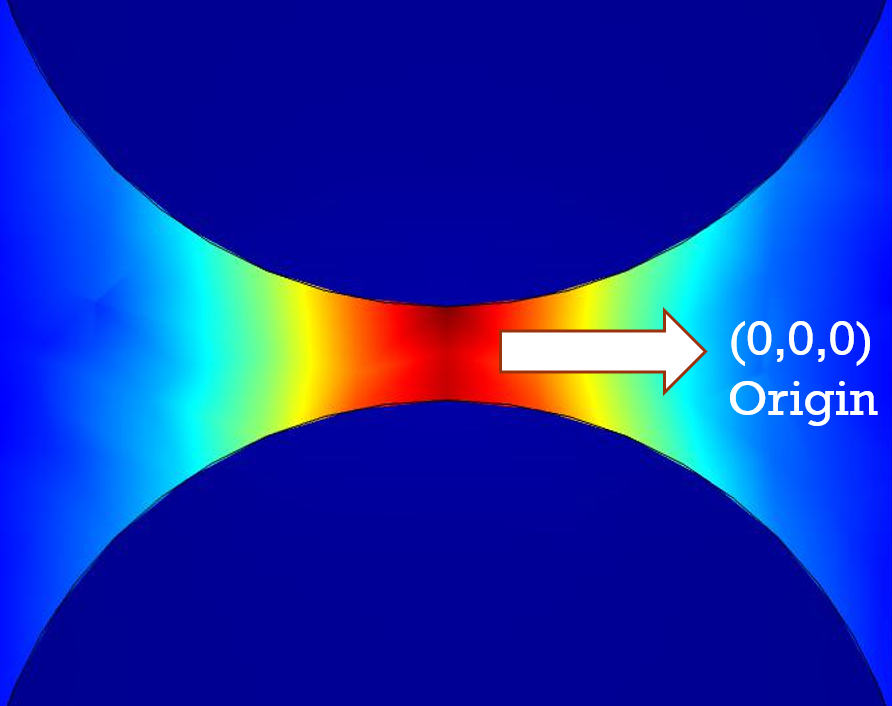}
\caption{Total electric field is plotted which shows the enhancement in the gap. The LFIEF is defined at the origin.}
\label{Fig:LFIF}
\end{figure}

Since optical nanoantennas do not possess waveguide ports like RF and microwave ports, it is essential to examine the near-field in regions very close to the antenna right from the beginning. In fact, the decision that a nanoantenna has been actually excited will be made based on some near-field measures like local near-field enhancement (see below.)
The near-field quantities we will be concerned with here include Local Field Intensity Enhancement Factor (LFIEF) and near-field scattered power (in the forward and backward directions). We start with LFIEF, which can be defined by the formula
\begin{equation}\label{LFIEF}
\text{LFIEF} :=  \frac{|\textbf{E}_t(\textbf{r}_c)|^2}{|\textbf{E}_b(\textbf{r}_c)|^2},
\end{equation}
where $|\textbf{E}_t(\textbf{r}_c)|$ is the amplitude of the total electric field at the center $\textbf{r}_c$ of the nanodipole, which is also the origin of the model as shown in Fig. \ref{Fig:LFIF}. In \eqref{LFIEF}, $|\textbf{E}_b(\textbf{r}_c)|$ is the amplitude of the background electric field at the center, which equals $0.36$ V/m in our case.

The electromagnetic power loss density $Q_e$ is equivalent to the standard average resistive (ohmic) losses in our model (calculated based on the fields inside the nanodipole).
The time average of the power outflow of the scattered fields is captured by $\textbf{S}_{{\rm sc}}$, i.e., the net time-average of the Poynting vector associated with the scattered fields.
Therefore, $Q_e$ and $\textbf{S}_{{\rm sc}}$ are expressions for the density of absorbed and scattered power, respectively. To find the absorbed and scattered power, we need to perform integration over the proper extensive domains. The absorbed power can be found by numerically integrating $Q_e$ over a quarter of the nanodipole mesh (by symmetry, see above) and then multiplying by 4, that is 
\begin{equation}
\label{eq:Ab}
P_{\rm abs}=4\int_{\frac{1}{4}\text{-dipole}} Q_e \mathop{\text{d}^3r}. 
\end{equation}
Total scattered power by the nanodipole $P_{\rm sc}$ can be found by integrating $\textbf{S}_{{\rm sc}}$ over the exterior surface of the quarter of the nanodipole and then multiplying by 4:
\begin{equation}
\label{eq:Sc}
P_{\rm sc}=4\oiint_{\frac{1}{4}\text{-}S} \textbf{S}_{{\rm sc}} \cdot  \mathop{\text{d}\textbf{s}},
\end{equation}
where $\text{d}\textbf{s}$ is the normal vector differential area pointing outwards from the exterior surface $S$ of the nanoantenna. 

\begin{figure}
\centering
\includegraphics[width=0.2125\textwidth]{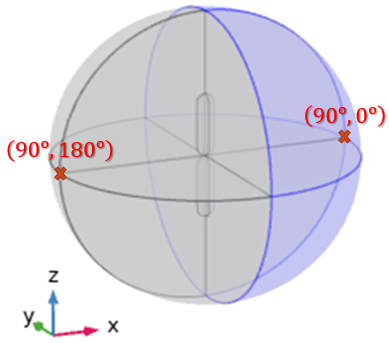}
\caption{Forward surface $S_f$ (blue to the right) and backward surface $S_b$ (grey to the left) corresponding to forward quantities and backward quantities integration, respectively.}
\label{Fig:forward}
\end{figure}

The power scattered in the near-field in the forward direction $P_{sc,f}^{\rm NF}$ can be calculated using \eqref{eq:Sc} by performing integration over the forward surface (the antenna surface that is towards the blue surface in Fig. \ref{Fig:forward}). On the other hand, the power scattered in the near-field in the backward direction $P_{sc,b}^{\rm NF}$ can be calculated by integrating over the backward surface (the antenna surface that is opposite to the blue surface in Fig. \ref{Fig:forward}). Note that symmetry can be exploited in the aforementioned integrations.
Extinction power can now be easily calculated by summing \eqref{eq:Ab} and \eqref{eq:Sc}:
\begin{equation}
\label{eq:Ext}
P_{\rm ext}=P_{\rm abs}+P_{\rm sc}.
\end{equation}
After finding the power absorbed and scattered by the nanodipole, we can compute the Standard Radiation Efficiency defined by
\begin{equation}\label{eq:RadEff}
e_{\rm std}:=\frac{P_{\rm sc}}{P_{\rm sc}+P_{\rm abs}}.
\end{equation}
Note that $P_{\rm abs}$ is usually negligible in classical antennas (RF regime) while it is significant in nanoantennas \cite{seok2013engineering}.


Standard radiation efficiency expression conventionally used in scattering phenomena is not indicative of the functionality of our transmitting optical nanoantenna. In this paper, the functionality criterion of our nanoantenna, which is illuminated by a subwavelength laser power, is that it scatters a maximum portion of power to the forward direction (far-field) and scatters a minimum portion of power to the backward direction (far-field). In order to provide some quantitative evaluation of our design, we consider the two types of radiated power to the far-field, one in the forward direction (away from the Tx antenna toward the location of the receiver), and \textit{backward scattered power}, which is the power reflected back from the nanodipole. The forward scattered power $P_{sc,f}^{\rm FF}$ can be obtained by integrating over the forward hemisphere surface shown in Fig. \ref{Fig:forward}, while backward scattered power $P_{sc,b}^{\rm FF}$ can be calculated by integrating over the backward hemisphere surface shown in the same Figure. It is important to note that the surfaces shown in Fig. \ref{Fig:forward} are the limits of our simulation domain (just before the PML) which means that they do not necessarily exist in the far-field. However, we will integrate over them (with exploiting symmetry) to find the far-field quantities (total, forward and backward scattered power) by using a far-field variable \footnote{For more information visit this series on COMSOL BLOG:
'Multiscale Modeling in High-Frequency Electromagnetics'} after scaling it down to the surface of integration. Total power scattered in the far-field can be obtained as:
\begin{equation}
\label{eq:intel}
P_{\rm sc}=4\oiint_{\frac{1}{4}\text{-}{S_{f,b}}} \frac{1}{2Z_0} \left( \frac{|A^{\rm FF}|}{r_{\rm sph}} \right)^2   \mathop{dS},
\end{equation}
where $S_{f,b}$ is the total surface area (blue surface and grey surfaces shown in Fig. \ref{Fig:forward}), $Z_0$ is free space intrinsic impedance, $|A^{\rm FF}|$ is the far-field variable which is basically equivalent to the so-called `scattering amplitude' in physics and it is calculated using  the  Stratton-Chu  formula  that  utilizes  the  near-field information  on  the  boundary  between  the  air  and  the  PML, and $r_{\rm sph}$ is the radius of the free space simulation domain just before the PML (the radius of the sphere shown in Fig. \ref{Fig:forward}). It is worthy to note that no physical dimension is assigned to $r_{\rm sph}$ since it is a scaling factor. This is important to maintain the dimensions of the formula. In \eqref{eq:intel}, the scattering amplitude $|A^{\rm FF}|$ is scaled down to the surface of integration because it is equal to the electric field amplitude at $1 m$ while our simulation domain extends up to $r_{sph}$ only. Note that $P_{sc,f}^{\rm FF}$ and $P_{sc,b}^{\rm FF}$ can be computed using \eqref{eq:intel} by limiting the integration to the blue surface $S_f$ and the grey surface $S_b$ shown in Fig. \ref{Fig:forward}, respectively (symmetry can be exploited.)




Generally speaking, $P_{\rm sc,f}$ is a function of distance $d_0$ and we explicitly write $P_{\rm sc,f}(d_0)$, we then have 
\begin{equation}\label{Power check}
    P_{\rm sc,f}(d_0)+P_{\rm sc,b}(d_0)=P_{\rm rad}.
\end{equation}
The forward and backward power computed at distance $d_0$ are both functions of $d_0$ but together they add up to produce a position independent sum, namely the total radiated power $P_{\rm rad}$, which is the same whether in the NF or FF zones. The relation \eqref{Power check} was used in what follows as a kind of consistency check of the computational model of the nanoantenna. Note that in \eqref{eq:intel}, total scattered power in the far-field is calculated and it is equal to the total power scattered in the near-field calculated in \eqref{eq:Sc}. This is why they have the same notation. However, a small difference can result due to our numerical model. 
As mentioned previously, a suitable optical antenna radiation efficiency measure can be defined using the quantities introduced above. We define Directed Radiation Efficiency as:
\begin{equation}
\label{eq:d_e}
e_{\rm dir}=\frac{P_{\rm sc,f}^{\rm FF}}{P_{\rm rad}+P_{\rm abs}}.
\end{equation}
This expression will be used in what follows in order to obtain a reasonable estimation of how the Tx antenna system is handling power considerations. It partially resolves the problem of lack of an analog of RF transmission line with definite mismatch factor because of the absence of an excitation physical port in our case.


\begin{figure*}
    \centering
    \subfigure[Local Field Intensity Enhancement Factor LFIEF.]
    {
        \includegraphics[width=0.483\textwidth]{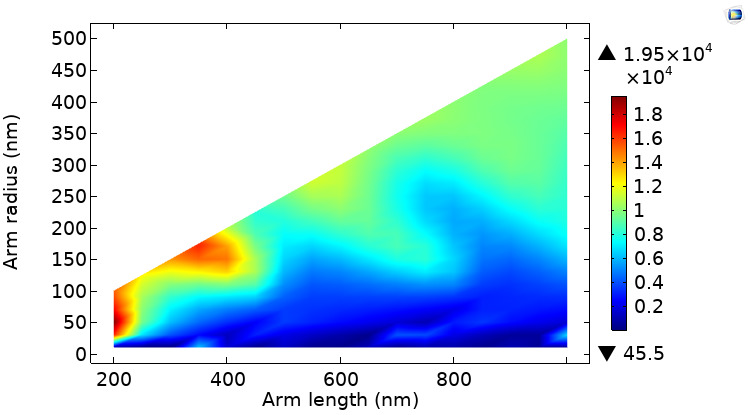}
        \label{Fig:LFIF2nm}
    }
    \subfigure[Total power scattered in near-field in all directions $P_{\rm sc}$.]
    {
        \includegraphics[width=0.483\textwidth]{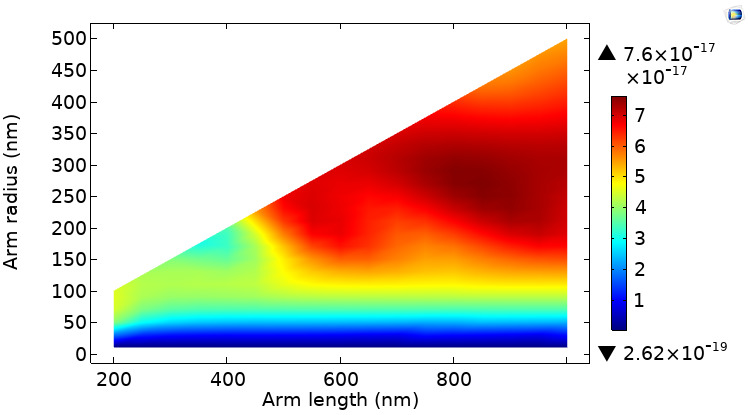}
        \label{Fig:Psca2nm}
    }
    \\
     \subfigure[Electromagnetic power (losses) in the nanodipole $P_{\rm abs}$.]
    {
        \includegraphics[width=0.483\textwidth]{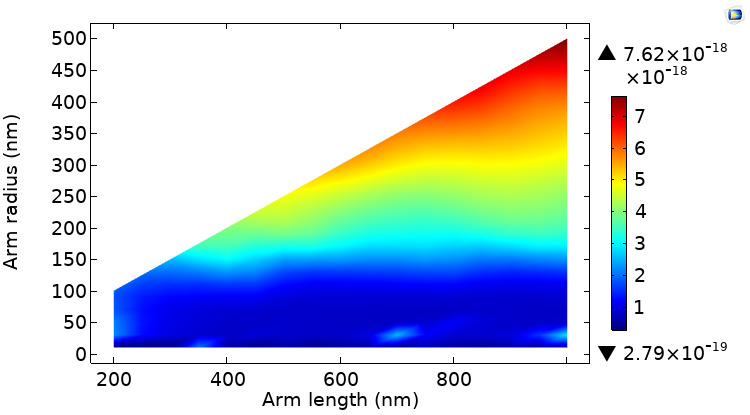}
        \label{Fig:Pabs2nm}
    }
    \subfigure[Extinction power of the nanodipole antenna $P_{\rm ext}$.]
    {
        \includegraphics[width=0.483\textwidth]{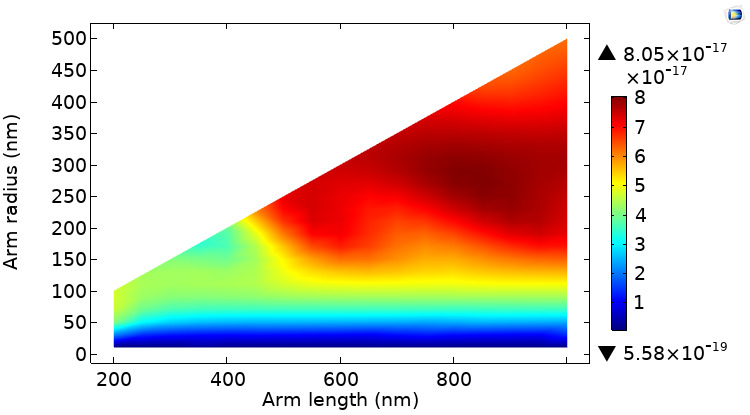}
        \label{Fig:Pext2nm}
    }
    \caption{Main near-field calculations.}
    \label{fig:NFFIGS}
\end{figure*}

 



In this paper, one of our objectives is to design an optical nanoantenna with a large directivity. Maximum directivity $D_{max}$ calculation will be performed. However, more information about the shape of the radiation pattern will be needed. Hence, directivity in forward and backward directions will be defined. We define forward directivity $D_f$ and backward directivity $D_b$ as 
\begin{equation}
\label{eq:Dfw}
D_{f}=\frac{|A^{\rm FF}(1,0,0)|^2}{{|A^{\rm FF}|^2}_{\rm avg}},
\end{equation}
\begin{equation}
\label{eq:Dbw}
D_{b}=\frac{|A^{\rm FF}(-1,0,0)|^2}{{|A^{\rm FF}|^2}_{avg}},
\end{equation}
respectively. Consequently, the Standard Gain and Directed Gain can be expressed by
\begin{equation}
\label{eq:std_gain}
G_{\rm std}=e_{\rm std}D_f,
\end{equation}
\begin{equation}
\label{eq:d_gain}
G_{\rm dir}=e_{\rm dir}D_f,
\end{equation}
respectively. These two measures are not identical and we propose that for complete evaluation of the performance of a Tx nanoantennas, both should be included. In a nutshell, the directive gain \eqref{eq:d_gain} allows the designer to estimate how much of the actually realizable gain is effectively usable for a receiving antenna placed in the forward direction. It does take into account what is the optical-antenna equivalent of \textit{reflection loss} $1-|\Gamma|^2$ in RF systems.

\section{Computational Analysis and Design Examples}
\label{sec:Computational Analysis and Design Examples}

The polarization of the incident field in addition to size, geometry and composition of the nanoantenna play an important role in the plasmonic effects and as a result in the optical properties \cite{csendur2009near}. In what follows, we will discuss the results of the parametric study corresponding to the near-field and far-field expressions discussed before in Section \ref{sec:Principal Models and Design Methodology}. 

\subsection{Near-Field Analysis}
\label{sec:NFR}

\begin{figure}
    \centering
    \subfigure[Scattered power in the near-field in forward and backward directions.]
    {
        \includegraphics[width=0.483\textwidth]{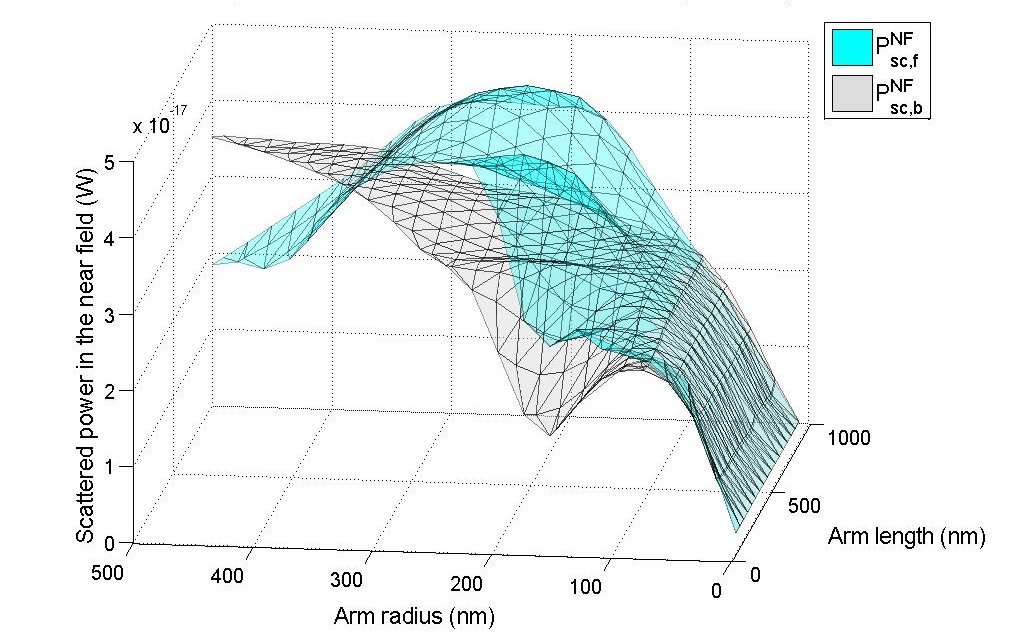}
        \label{Fig:Pscaf2nm}
    }
    \\
    \subfigure[Scattered power in the far-field in forward and backward directions.]
    {
        \includegraphics[width=0.483\textwidth]{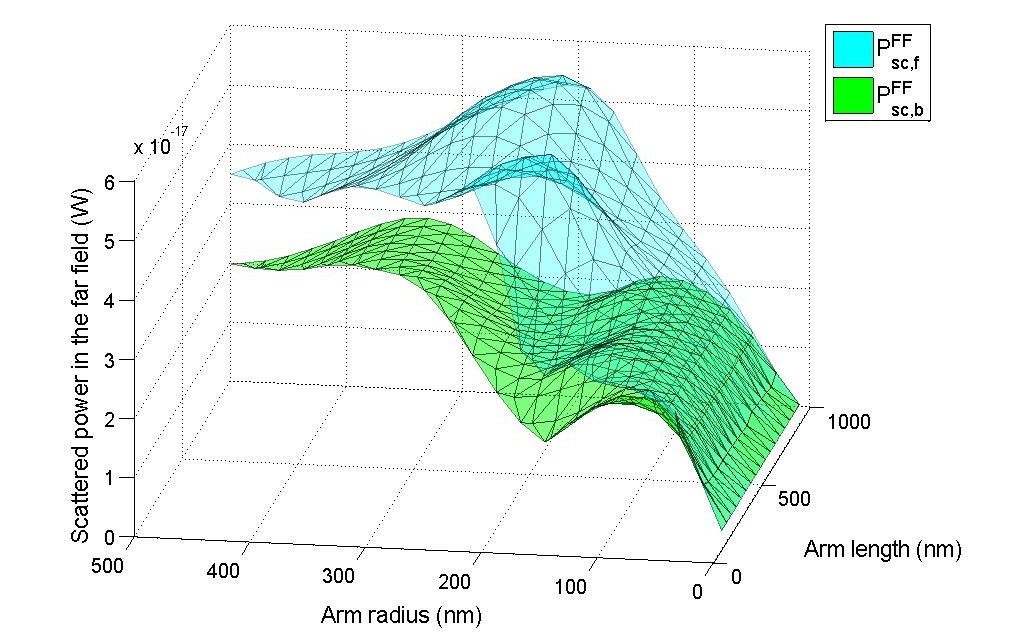}
        \label{Fig:Pscaff-f2nm}
    }
    \caption{Scattered power in the forward and backward directions in the near and far-field.}
    \label{fig:ScFBNFFF}
\end{figure}

The main near-field results calculated in any scattering problem are shown in Fig. \ref{fig:NFFIGS}. 
In Fig. \ref{Fig:LFIF2nm}, LFIEF is plotted versus arm length and arm radius. It has large values for most dimensions. The LFIEF generally increases with increasing radius for a fixed arm length. The peak values occur for small arm length i.e. when the laser is not focused compared to the dimensions. 
The total scattered power is shown in Fig. \ref{Fig:Psca2nm}. It is evident that scattering depends mainly on arm radius value since it is proportionally related to it. It does not generally depend on arm length. 
Here, the power absorbed by the nanoantenna $P_{\rm abs}$ will be considered in the near-field calculations. It is a part of the extinction formula. The absorbed power (electromagnetic losses) in the nanoantenna is shown in Fig. \ref{Fig:Pabs2nm}. The arm length has no effect here and the higher the radius, the higher the absorbed power. Adding up Fig. \ref{Fig:Psca2nm} and Fig. \ref{Fig:Pabs2nm} gives the extinction power $P_{\rm ext}$ shown in Fig. \ref{Fig:Pext2nm}. It is evident that the extinction is dominated mainly by scattering for large radii and by both absorption and scattering for very small radii. The reason for this is that absorption is generally smaller than scattering, except for very small radius values in which they are comparable. 
Fig. \ref{fig:ScFBNFFF} shows the scattered power in the forward and backward directions in the near and far-fields. Regarding the near-field scattered power in the forward direction $P_{sc,f}^{\rm NF}$ and that in the backward direction $P_{sc,b}^{\rm NF}$, they generally do not depend on the arm length as shown in Fig. \ref{Fig:Pscaf2nm}. They are both small and almost equal in small radii cases but afterwards, they increase monotonically with radius (with $P_{sc,f}^{\rm NF}$ having a larger value). This continues until $P_{sc,f}^{\rm NF}$ reaches its peak around 250 nm radius. After this peak, it decreases while $P_{sc,b}^{\rm NF}$ continues to increase with radius. To sum up, for centered radii values around 250 nm radius, the power is mostly scattered in the forward direction in contrast to the situation of small and large radii values. This near-field scattered power information is significant in case of utilizing the nano-antenna in near-field communications.

\subsection{Far-Field Analysis}

\begin{figure*}
    \centering
    \subfigure[Directivity in the forward direction $D_f$.]
    {
        \includegraphics[width=0.483\textwidth]{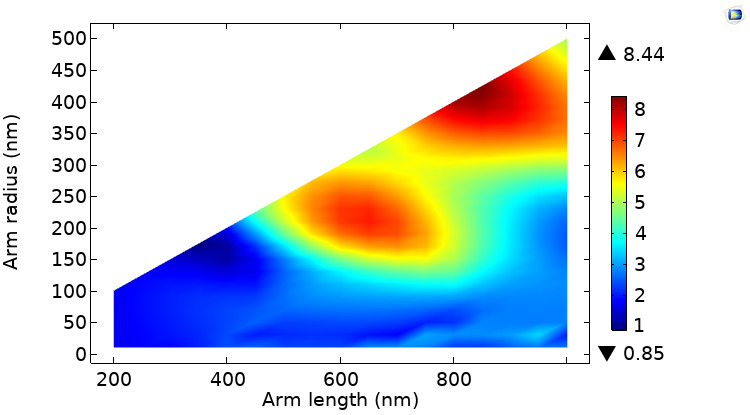}
        \label{Fig:Df2nm}
    }
    \subfigure[Directivity in the backward direction $D_b$.]
    {
        \includegraphics[width=0.483\textwidth]{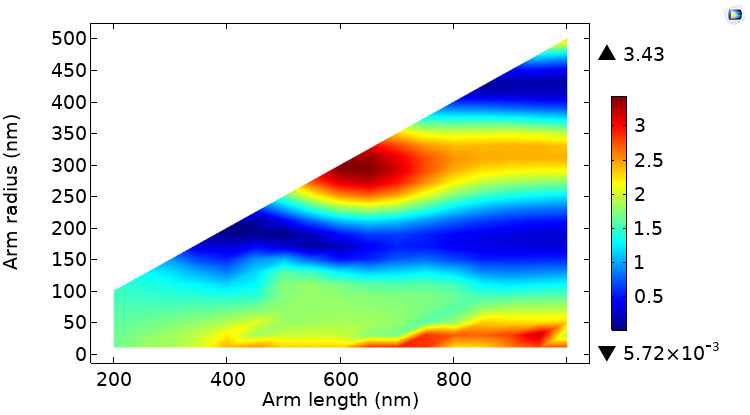}
        \label{Fig:Db2nm}
    }
    \\
     \subfigure[Difference between $D_f$ and $D_b$.]
    {
        \includegraphics[width=0.483\textwidth]{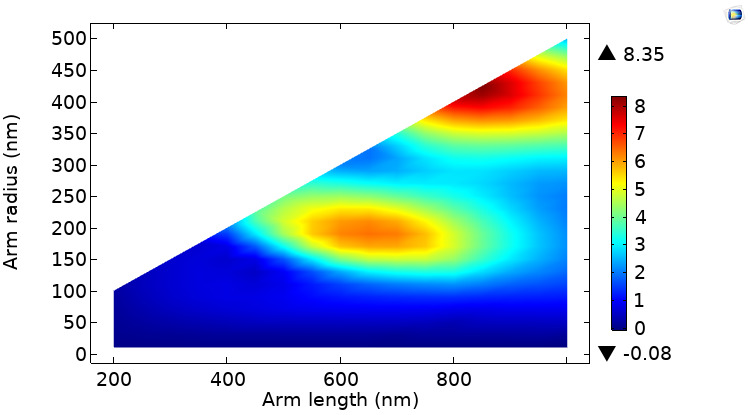}
        \label{Fig:DfmDb2nm}
    }
    \subfigure[Difference between $D_{max}$ $D_f$.]
    {
        \includegraphics[width=0.483\textwidth]{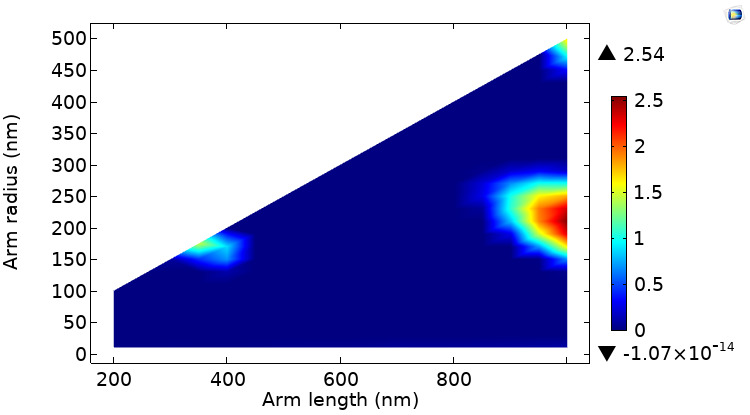}
        \label{Fig:DmDf2nm}
    }
    \caption{Directivity calculations and their relations.}
    \label{fig:Abstraction}
\end{figure*}

\begin{figure}
\centering
\includegraphics[width=0.4\textwidth]{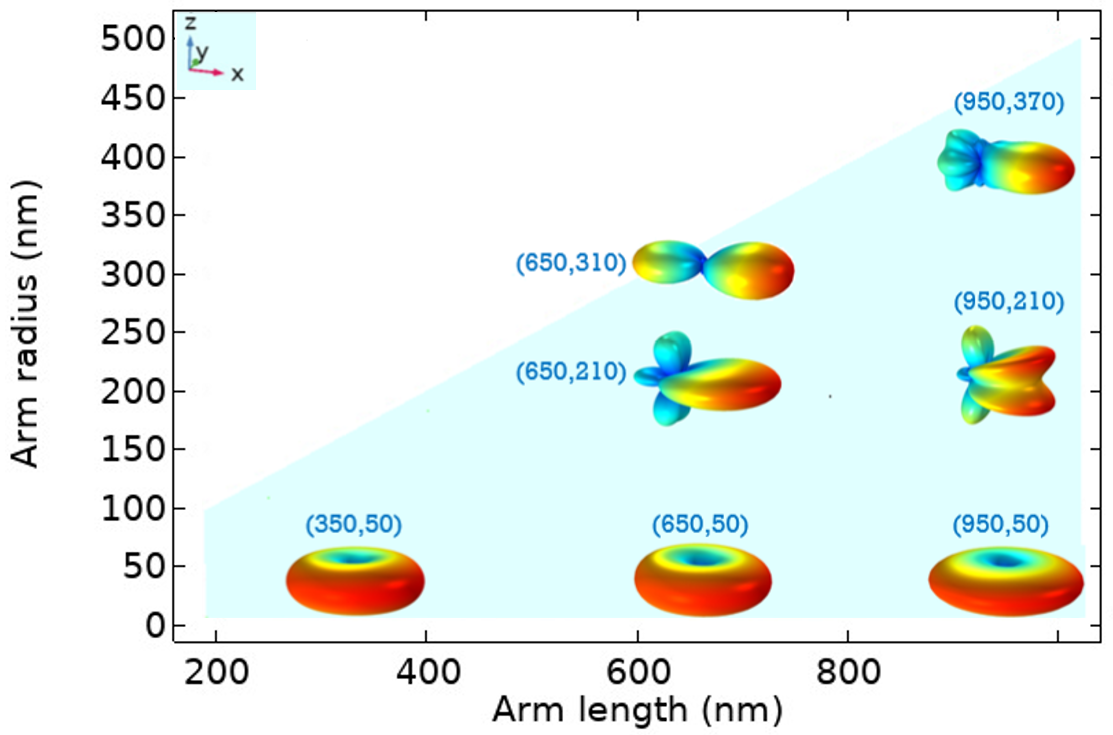}
\caption{Radiation patterns of the indicated dimensions.}
\label{Fig:nice}
\end{figure}

\begin{figure*}
    \centering
    \subfigure[Standard radiation efficiency $e_{\rm std}$.]
    {
        \includegraphics[width=0.483\textwidth]{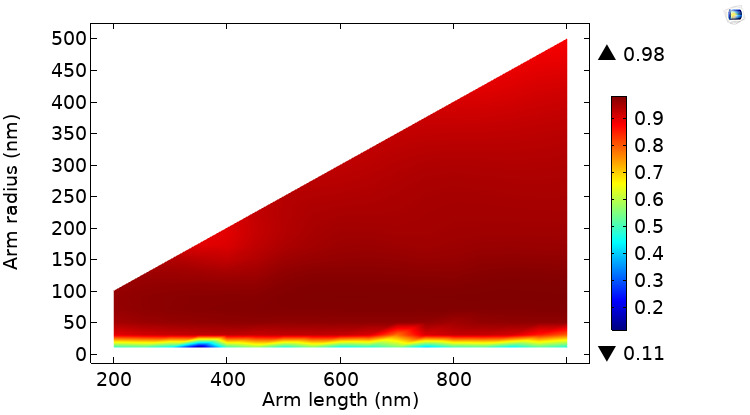}
        \label{Fig:er2nm}
    }
    \subfigure[Directed efficiency $e_{\rm dir}$.]
    {
        \includegraphics[width=0.483\textwidth]{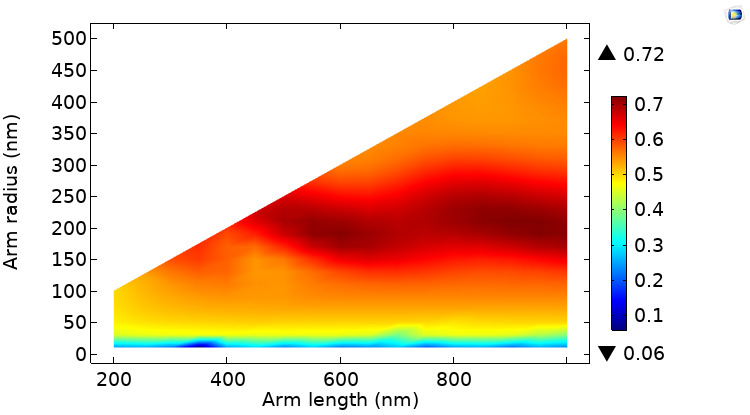}
        \label{Fig:deff2nm}
    }
    \\
     \subfigure[Standard gain $G_{\rm std}$.]
    {
        \includegraphics[width=0.483\textwidth]{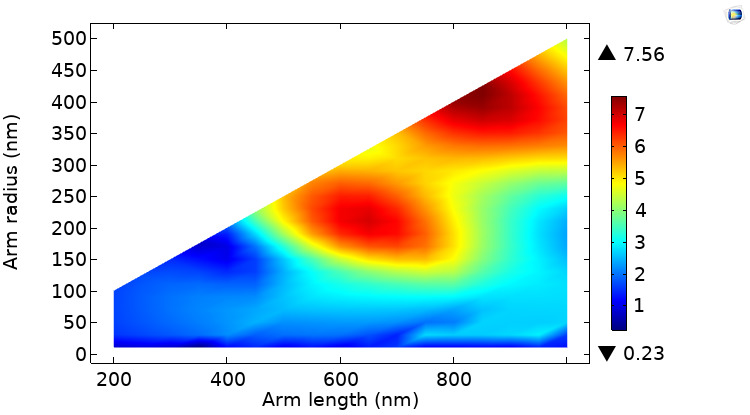}
        \label{Fig:stdgain2nm}
    }
    \subfigure[Directed gain in forward direction $G_{\rm dir}$.]
    {
        \includegraphics[width=0.483\textwidth]{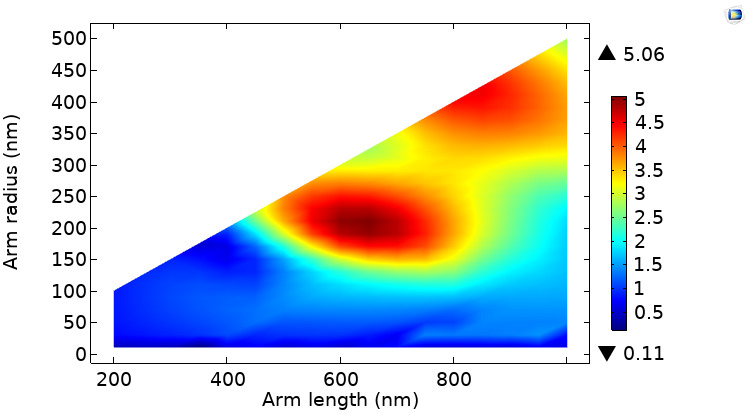}
        \label{Fig:dgain2nm}
    }
    \caption{Main far-field calculations.}
    \label{fig:FFFIGS}
\end{figure*}

The far-field mainly depend on the $A^{\rm FF}$ variable, here computed in our model via calculated using the Stratton-Chu formula. As discussed in Section \ref{sec:Near-Field and Far-Field Calculations}, the distribution and density of scattered power in the far-field will determine the main design parameters we are concerned with: Directivity and Gain. 
The far-field scattered power in forward direction $P_{sc,f}^{\rm FF}$ and backward direction $P_{sc,b}^{\rm FF}$ are plotted in Fig. \ref{Fig:Pscaff-f2nm}. The strong dependence on arm radius also holds and $P_{sc,f}^{\rm FF}$ is generally larger than $P_{sc,b}^{\rm FF}$ for all cases except for small arm radii cases for which $P_{sc,f}^{\rm FF}$ and $P_{sc,b}^{\rm FF}$ are relatively small and almost equal. The largest directed power (difference between $P_{sc,f}^{\rm FF}$ and $P_{sc,b}^{\rm FF}$)  occurs for the radii between 150 nm and 230 nm. This information helps to understand the Directed Efficiency distribution  (see below). 
Note that the four quantities $P_{sc,f}^{\rm NF}$, $P_{sc,b}^{\rm NF}$, $P_{sc,f}^{\rm FF}$ and $P_{sc,b}^{\rm FF}$ shown in Fig. \ref{fig:ScFBNFFF} are governed by \eqref{Power check} for any selected nanoantenna dimension. The four quantities are almost equal for very small radii values.
Despite a very small difference due to numerical calculations, the total power scattered in the far-field is almost equal to that in the near-field $P_{\rm sc}$ which is shown in Fig. \ref{Fig:Psca2nm}. The same comments also hold here; the scattered power depends generally on the radius value, not on the arm length. As mentioned before, the scattered power in the near-field is evaluated by integrating the scattered power over the surface of the nanoantenna while the power scattered in far-field is evaluated using the far-field variable which is calculated using the Stratton-Chu formula that utilizes the near-field information on the boundary between the air and the PML. The fact that the two calculations are equivalent gives a strong indication of the accuracy and validity of our simulation model. The small difference between them can be made even smaller by building a more accurate mesh but with the cost of increasing the computational burden.

Directivity calculations are shown in Fig. \ref{fig:Abstraction}. The forward directivity $D_f$ results are shown in Fig. \ref{Fig:Df2nm}. It is clear that both arm length and radius affect $D_f$. The peak (8.44 = 9.3 dB) occurs for the nano-sphere case with 425 nm radius nano-spheres. The second highest directivity peak (7.3 = 8.6 dB) occurs around 650 nm arm length and 210 nm arm radius. Backward directivity $D_b$ is much smaller than $D_f$ and its peak (3.43 = 5.35 dB) occurs at the two nano-spheres case with 300 nm radius as shown in Fig. \ref{Fig:Db2nm}. The result of subtracting $D_b$ from $D_f$ is shown in Fig. \ref{Fig:DfmDb2nm} which indicates that $D_f$ is always higher than or equal to $D_b$ (the negative values are ignored since they are very small). In this work, we consider $D_f$ for our design calculations since it is often equal to $D_{\rm max}$. For the few cases where $D_f$ is not equal to $D_{\rm max}$, then these cases have a directivity that is not significant so there is no need to consider them in our search for an optimum nanoantenna as noted from Fig. \ref{Fig:DmDf2nm} which shows the result of subtracting $D_{\rm max}$ from $D_{f}$. These few cases occur mostly around 1000 nm arm length and 210 nm radius. In fact, these calculations shown in Fig. \ref{fig:Abstraction} help to introduce numerical description to the radiation pattern shapes. This multi-aspect numerical description reveals the main regions of the main radiation pattern shapes which transit gradually between the regions. Fig. \ref{Fig:nice} shows these main radiation pattern shapes. The cases considered have the following coordinates: (350 nm, 50 nm), (650 nm, 50 nm), (950 nm, 50 nm), (650 nm, 210 nm), (950 nm, 210 nm), (650 nm,310 nm), (950 nm, 370 nm). Axis orientation is fixed for all radiation pattern shapes and is shown in the upper left corner of the figure. For the few cases where forward directivity is not maximum, then the radiation pattern looks like a horn (two peaks around the forward direction) as shown in the (950 nm, 210 nm) case in Fig. \ref{Fig:nice}. Note that all radiation patterns are symmetric as was previously discussed. 

The Standard Radiation Efficiency results $e_{\rm std}$ are shown in Fig. \ref{Fig:er2nm}. They are very large for all cases except for the very small arm radius value (10 nm). The reason behind this is that absorption at these values is comparable to scattering as indicated in Section \ref{sec:NFR}. The radius values between 50 nm and 110 nm exhibit the largest values that are close to unity. It is evident that the arm length generally almost has no effect here and this comes from the fact that the $P_{\rm sc}$ and $P_{\rm abs}$ do not depend on the arm length as was previously explained. On the other hand, the Directed Radiation Efficiency $e_{\rm dir}$ peak exists around 200 nm radius with a value of 72\% as shown in Fig. \ref{Fig:deff2nm}. The previously mentioned 650 nm arm length and 210 nm radius antenna is included in this region with a $e_{\rm dir}$ of 69\%. 
The Standard Gain $G_{\rm std}$ results are shown in Fig. \ref{Fig:stdgain2nm} with a maximum of (7.56 = 8.8 dB) at two nano-sphere case (radius of 410 nm for each). The second-largest $G_{\rm std}$ peak (6.94 = 8.4 dB) occurs at 650 nm arm length and 210 nm radius. The Directed Gain $G_{\rm dir}$ results, which are our main concern here, are shown in Fig. \ref{Fig:dgain2nm}. They have a maximum of (5.06 = 7 dB) at the dimensions (650 nm arm length and 210 nm radius) which is a strong indication of the optimality of these dimensions in all respects. To sum up, the parameters of the optimum nano-antenna are summarized in Table \ref{table1}.

\begin{table}
\centering
\caption{Optimum nanoantenna parameters}
\label{table1}
\begin{tabular}{|*{12}{p{2cm}|}}
\hline
Parameter       & Value \\ \hline
Arm length     & $650$ nm  \\ \hline
Arm radius   & $210$ nm  \\ \hline
Gap        & $2$ nm        \\ \hline
$e_{\rm std}$       & $0.95$       \\ \hline
$e_{\rm dir}$        & $0.69$       \\ \hline
$D_f$       & $8.6$ dB     \\ \hline
$G_{\rm std}$        & $8.4$ dB      \\ \hline
$G_{\rm dir}$       & $7$ dB      \\ \hline
\end{tabular}
\end{table}

\subsection{Polarization diversity in Optical Antennas}
\label{sec:Optical Communication system using polarization diversity}

\begin{figure}
\centering
\includegraphics[width=0.5\textwidth]{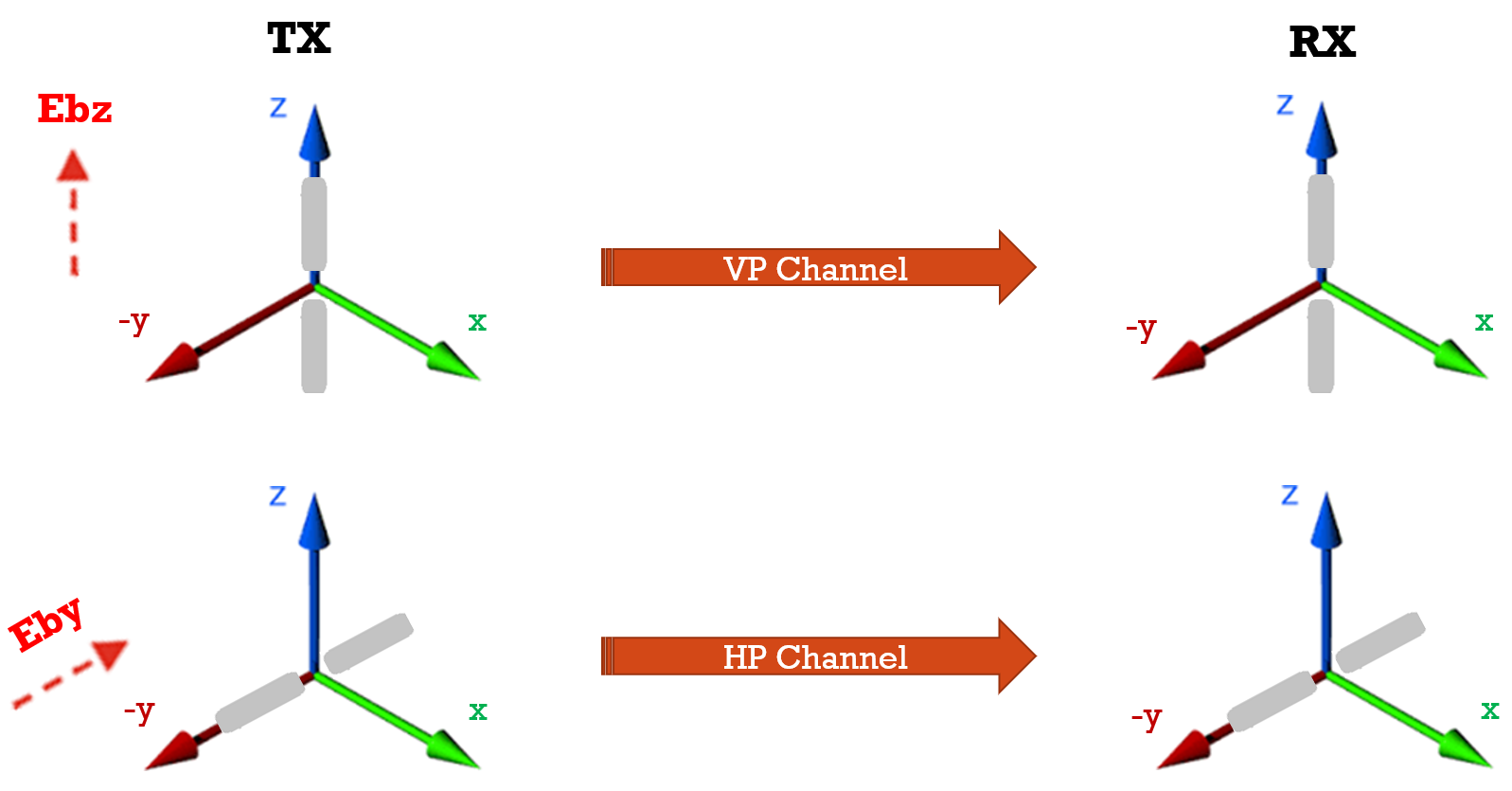}
\caption{Polarization diversity design consideration for optical antenna communications.}
\label{Fig:polarizationd}
\end{figure}

In order to achieve a communication system using the original nanoantenna transmitter proposed in this work, the polarization diversity technique is used. Each side of our communication system has two nanoantennas as shown in Fig. \ref{Fig:polarizationd}. One of them is on the z-axis (Vertical Polarization Channel) and the other is on the y-axis (Horizontal Polarization Channel). The implementation of our proposed nanoantenna using polarization diversity technique in the nano-scale is achieved by monitoring LFIEF at the gaps of the two nanoantennas in the Rx side.
Initially, let's consider only the fields generated from the Tx z-axis nanodipole when it is illuminated by a subwavelength laser polarized in the z-direction. In the far-field, the generated field will have a $\theta$- and a $\phi$-components. At $1$m distance, they are equal to $A_{\theta}^{\rm FF}$ and $A_{\phi}^{\rm FF}$, respectively\footnote{Note that the far-field zone starts in shorter distance than $1m$ but here we will assume the distance is $1$m in order to use the scattering amplitude $A^{\rm FF}$ which corresponds to the field at $1$m. The electric field distribution at any distance in the far-field zone is analogous to what will be shown next with the exception that it is multiplied by a linear scaling factor which will make it larger and able to excite the Rx side unlike the fields that will be shown next (which will only help to decide the optimum nano-antenna). Also, exciting subwavelength laser amplitude can be made much larger to ensure that the Rx is excited.}. $|A_{\theta}^{\rm FF}|$ generated is along the major axis of the z-axis Rx nanodipole and this will result in high LFIEF in its gap. However, this won't result in a significant field enhancement in the y-axis Rx nanoantenna gap since $|A_{\theta}^{\rm FF}|$ is perpendicular to the major axis. On the other hand, $|A_{\phi}^{\rm FF}|$ generated from the z-axis Tx nanodipole, assuming it is large enough, will result in LFIEF in the gap of the y-axis Rx nanodipole. However, it won't result in significant LFIEF in the z-axis Rx nanodipole gap. Consequently, $|A_{\theta}^{\rm FF}|$ generated from the z-axis Tx nanoantenna must be designed to be high enough so that it results in an enhancement in the z-axis Rx nanoantenna gap and $|A_{\phi}^{\rm FF}|$ must be small to reduce the enhancement in the y-axis Rx nanoantenna (which is related to the horizontal channel), i.e. this will eliminate the error of `vertical polarization channel is detected given that horizontal channel is the active one'.
By analogy, the y-axis Tx nanodipole that is excited by a y-polarized subwavelength laser is responsible for the horizontal channel and it must be designed to generate high $|A_{\phi}^{\rm FF}|$ to cause an enhancement in the gap of the Rx y-axis nanoantenna. It must also be designed to generate small $|A_{\theta}^{\rm FF}|$ to reduce field enhancement on the z-axis Rx nanoantenna (which is related to the vertical channel), i.e. this will eliminate the error of `horizontal polarization channel is detected given that vertical channel is the active one'.

\begin{figure}
\centering
\includegraphics[width=0.5\textwidth]{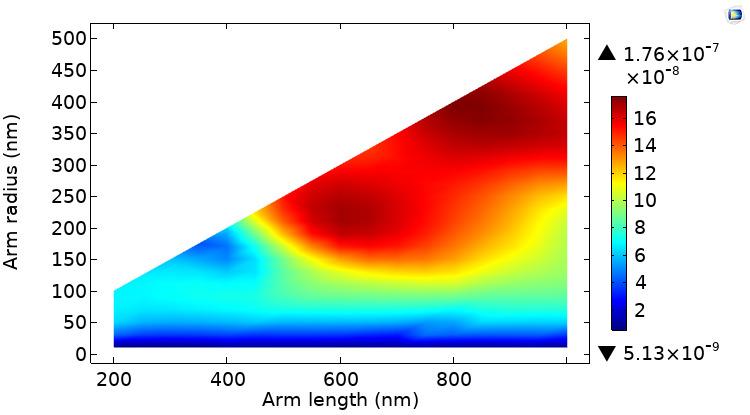}
\caption{The results of $|A_{\theta}^{\rm FF}({90^{^{\circ}}, 0^{^{\circ}}})|$.}
\label{Fig:Efartheta2nm}
\end{figure}
\begin{figure}
\centering
\includegraphics[width=0.5\textwidth]{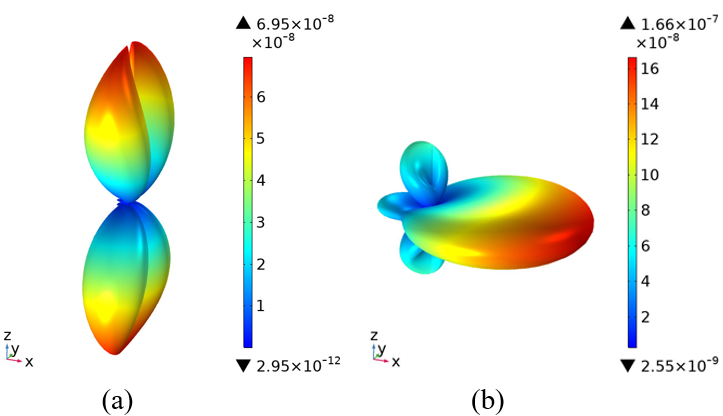}
\caption{Radiation pattern of (a) $|A_{\phi}|$ and (b) $|A_{\theta}|$ generated from the 650 nm arm length and 210 nm radius nanodipole.}
\label{Fig:Efarphitheta}
\end{figure}
\begin{figure}
\centering
\includegraphics[width=0.49\textwidth]{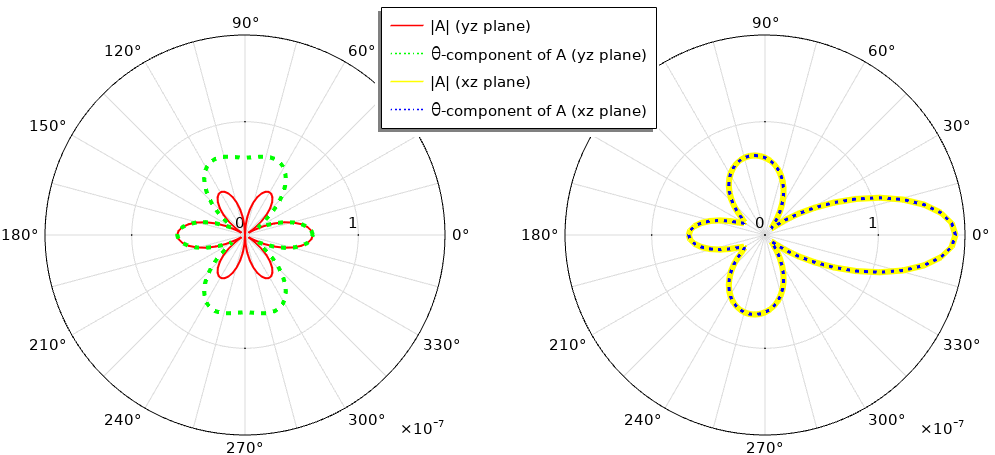}
\caption{Radiation patterns of $|A|$ and $|A_{\theta}|$ in the xy and yz planes. Both are plotted for the same scale.}
\label{Fig:xyxz}
\end{figure}

In fact, the two nanodipoles in the Tx side are exactly the same since the optimized z-axis Tx nanoantenna that gives best electric field in the vertically polarized channel will have the same dimensions as that which gives the best electric field in the horizontally polarized channel since it is basically the same operating nanoantenna but rotated by $90^{^{\circ}}$ and its corresponding excitation is also rotated with it. Therefore, our problem now is to optimize the z-axis Tx nanodipole for maximum $|A_{\theta}^{\rm FF}|$ in the forward direction and hence will have to optimize $|A_{\theta}^{\rm FF}(\theta,\phi)|$ at $\theta = 90^{^{\circ}}$, $\phi = 0^{^{\circ}}$. 
Computations of $|A_{\theta}^{\rm FF}({90^{^{\circ}}, 0^{^{\circ}}})|$ are shown in Fig. \ref{Fig:Efartheta2nm}. The distribution of the peaks is quite similar to the directed gain shown in Fig. \ref{Fig:dgain2nm}. The optimum dimensions occur for two nano-sphere antenna with a 400 nm radius for each which gives a value of $1.76*10^{-7}$ V/m. The previously optimized nanodipole for maximum directed gain (165 nm, 210 nm) gives a value of $1.68*10^{-7}$ V/m which is very close to the major peak and it is almost the peak for the nanodipole case. 
Radiation pattern of $|A_{\phi}|$ and $|A_{\theta}|$ generated from the 650 nm arm length and 210 nm radius nanodipole are shown in Fig. \ref{Fig:Efarphitheta}. The $\theta$-component is very directive in the forward direction. The $\phi$-component gives zero electric field in the forward direction which will eliminate errors in reception. $|A_{\phi}^{\rm FF}|$ has smaller peak values which indicate small power is lost in the undesired polarization. The radiation patterns of $|A|$ and $|A_{\theta}|$ in the xy and yz planes are shown in Fig. \ref{Fig:xyxz}. Note that $|A|$ is identical to $|A_{\theta}|$ in the xz plane. 

\section{Conclusion}
\label{sec:Conclusion and Future Work}

In this paper, a transmitting nanoantenna was proposed using a subwavelength laser
excitation mechanism at 800 nm wavelength with 200 nm spot radius. The nanoantenna model was designed using an efficient full-wave electromagnetic multiphysics software based on Finite Element Method. The nanoantenna design stages were discussed in detail, including geometry, material composition, discretization mesh, physics, and choice of suitable quantitative performance measures. Two geometries were considered for the nanoantenna; nanodipole and two nano-spheres separated by a gap. Also, post-processing expressions that were manually entered to monitor the near-field and far-field quantities were explained. The nanoantenna was optimized by performing a parametric sweep over radius and length. The radiation pattern changed with changing the dimensions, varying between omni-directional, directed and horn radiation patterns. An optimum nanoantenna in terms of directed gain (7 dB) was achieved with 650 nm arm length and 210 radius. Furthermore, the polarization diversity aspects of the optical antenna system were investigated and some design data were given. 
This work can be extended by investigating other transmit geometries,
such as single nano-sphere, single nano-rod, group of spheres, group of nanorods, bowtie and
Yagi-Uda. Also, the gap between the two NPs can be swept in order to understand it’s effect
on the results. Moreover, sub-wavelength laser wavelength, spot radius, and polarization type
can be further optimized to maximize their effects on the nanoantenna radiation. 

\section*{Acknowledgment}

The authors acknowledge the financial support of the Deanship of Scientific Research at the Jordan University of Science and Technology under research grant number 2017/399. The authors are also thankful to Dr. Andrew  Strikwerda for providing helpful feeding on our the computational model.


%



\ifCLASSOPTIONcaptionsoff
  \newpage
\fi



%




\bibliographystyle{IEEEtran}

\end{document}